\documentclass[aps,prl,twocolumn,groupedaddress]{revtex4-1}
\usepackage{graphicx}

\begin{document}


\title{Constraints on annihilating dark matter in the Omega Centauri cluster}


\author{Man Ho Chan, Chak Man Lee}
\affiliation{Department of Science and Environmental Studies, The Education University of Hong Kong \\ 
Tai Po, New Territories, Hong Kong, China}


\begin{abstract}
Recent gamma-ray and radio studies have obtained some stringent constraints on annihilating dark matter properties. However, only a few studies have focussed on using X-ray data to constrain annihilating dark matter. In this article, we perform the X-ray analysis of annihilating dark matter using the data of the Omega Centauri cluster. If dark matter is the correct interpretation of the non-luminous mass component derived in the Omega Centauri cluster, the conservative lower limits of thermal dark matter mass annihilating via the $\tau^+\tau^-$, $b\bar{b}$ and $W^+W^-$ channels can be significantly improved to 104(43) GeV, 650(167) GeV and 480(137) GeV respectively, assuming the diffusion coefficient $D_0 \le 10^{26}(10^{27})$ cm$^2$/s. These constraints can safely rule out the recent claims of dark matter interpretation of the gamma-ray excess and anti-proton excess seen in our Galaxy. Generally speaking, the conservative lower limits obtained for non-leptophilic annihilation channels are much more stringent than that obtained by gamma-ray analysis of nearby dwarf galaxies. We anticipate that this would open a new window for constraining annihilating dark matter.
\end{abstract}

\maketitle


\section{Introduction}
Recently, multi-messenger studies of galaxies and galaxy clusters have obtained many stringent constraints of annihilating dark matter, including the gamma-ray studies of our Galaxy and nearby dwarf galaxies \cite{Ackermann2,Ackermann,Albert,Boddy,Li}, radio studies of galaxies \cite{Egorov,Chan,Kar,Chan6,Vollmann2,Regis2} and galaxy clusters \cite{Colafrancesco,Storm,Colafrancesco2,Chan3,Chan4}, cosmic-ray studies of our Galaxy \cite{Ambrosi,Aguilar}, and neutrino studies of our Galaxy \cite{Antares}. Different kinds of studies might have different advantages and disadvantages. For example, for radio studies, although the sensitivity and resolution of radio observations can be very high, the uncertain magnetic field strength contributes the largest systematic errors in the analysis. For neutrino studies, the observational uncertainties of astrophysical neutrino flux are usually quite large. Also, the sensitivity of neutrino detection is not good enough to get stringent constraints \cite{Antares}. In comparison, the constraints obtained from gamma-ray studies are generally more stringent and more robust because the involved systematic uncertainties are usually smaller. The only uncertainties are the dark matter density profile, the background point source contribution due to pulsars, and the diffuse backgrounds in modeling the regions around point sources. Another drawback is that current resolution of gamma-ray detection is larger than $5'$ so that more distant galaxies and galaxy clusters with small angular sizes ($<1'$) are difficult to be analyzed in gamma-ray studies to constrain dark matter.

In view of the multi-messenger studies, there are only a few studies using X-ray data to constrain annihilating dark matter. Many previous multi-wavelength studies of dark matter have included radio and gamma rays, but not X-ray \cite{McDaniel,Bhattacharjee}. Most of the related X-ray studies are focussing on the decaying or leptophilic annihilating dark matter signals due to keV dark matter (e.g. keV sterile neutrinos) \cite{Bulbul,Boyarsky} or light dark matter (MeV or sub-GeV) \cite{Essig,Laha,Cirelli2}. In fact, the inverse Compton scattering (ICS) of Cosmic Microwave Background (CMB) radiation due to the electrons and positrons produced from GeV to sub-TeV dark matter annihilation can boost the CMB photon energy to X-ray bands. Therefore, it has been suggested for a long time that using X-ray data of dwarf galaxies can give certain constraints on annihilating dark matter with mass $>$ GeV \cite{Jeltema,Jeltema2,McDaniel2,Cirelli2}. However, only a few studies have put constraints on GeV to sub-TeV annihilating dark matter using X-ray data and the constraints obtained are not very stringent \cite{Colafrancesco3,Regis,Chan5}. 

Using X-ray data of dwarf galaxies to constrain dark matter is a very good option because the physics of ICS is well-known and the CMB spectrum is well-determined. Also, the amount of hot gas in dwarf galaxies is very small so that the background diffuse X-ray contribution is not significant. This would greatly suppress the unwanted background X-ray flux so that we can get more stringent constraints on the annihilating dark matter parameters. In this article, we show that using the X-ray data of the Omega Centauri cluster ($\omega$-Cen) can give very stringent constraints on annihilating dark matter. The lower limits of dark matter mass can be more stringent than that obtained in gamma-ray studies of dwarf galaxies. Note that the existence of dark matter in $\omega$-Cen is still not 100\% confirmed. Although the non-luminous mass component in $\omega$-Cen is strongly statistically preferred relative to a stellar mass-only model based on recent kinematic studies \cite{Brown,Evans}, the lower end of the allowed dynamical mass range is still plausibly consistent with the mass contained in stellar remnants. Nevertheless, if dark matter is the correct interpretation of the non-luminous mass component, our analysis would give exciting new constraints of annihilating dark matter from X-ray. We anticipate that this would open a new window for constraining annihilating dark matter.

\section{The X-ray analysis of annihilating dark matter}
A large amount of high-energy electrons and positrons would be produced from dark matter annihilation. The diffusion and cooling of these electrons and positrons can be governed by the diffusion-cooling equation \cite{Atoyan}
\begin{equation}
\frac{\partial}{\partial t}\frac{dn_{\rm e}}{dE}=\frac{D(E)}{r^2}\frac{\partial}{\partial r}\left(r^2\frac{\partial }{\partial r}\frac{dn_{\rm e}}{dE}\right)+\frac{\partial}{\partial E}\left[b(E)\frac{dn_{\rm e}}{dE}\right]+Q(E,r),
\label{diffusion}
\end{equation}
where $dn_{\rm e}/dE$ is the electron/positron density spectrum, $D(E)$ is the diffusion function, $b(E)$ is the cooling function, and $Q(E,r)$ is the source density spectrum from dark matter annihilation. The diffusion function is usually written in terms of an energy-dependent function $D(E)=D_0(E/1\;{\rm GeV})^{\delta}$, where $D_0$ is the diffusion coefficient and $\delta$ is the diffusion index. The source density spectrum is given by
\begin{equation}
Q(E,r)=\frac{\langle \sigma v \rangle [\rho_{\rm DM}(r)]^2}{2m_{\rm DM}^2}\frac{dN_{\rm e,inj}}{dE},
\end{equation}
where $m_{\rm DM}$ is the dark matter mass, $\rho_{\rm DM}(r)$ is the dark matter density profile and $dN_{\rm e,inj}/dE$ is the injection energy spectrum of dark matter annihilation, which can be calculated theoretically for different annihilation channels \cite{Cirelli}. In our first analysis, we will set the annihilation cross section $\langle \sigma v \rangle$ to be a free parameter and get its upper bound. Then in the second part, we will specifically consider the thermal annihilation cross section and get the constraints on dark matter mass.

In equilibrium, the diffusion-cooling equation can be solved by the Green's function method \cite{Vollmann}:
\begin{equation}
\frac{dn_{\rm e}}{dE}=\frac{1}{b(E)} \int dE' \int dr' \frac{r'}{r}G(E,E',r,r')Q(E',r'),
\end{equation} 
where the Green's function can be written in terms of a series of Fourier functions and the ratio $b(E')/D(E')$ as \cite{Vollmann}:
\begin{eqnarray}
G(E,E',r,r')&=&\frac{2}{r_h} \sum_{j=1}^{\infty} \sin \left(\frac{j \pi r'}{r_{\rm h}} \right) \sin \left(\frac{j \pi r}{r_{\rm h}} \right) \nonumber\\
&&\times \left[\frac{r_{\rm h}^2b(E')}{j^2\pi^2D(E')} \delta(E-E') \right.\nonumber\\
&& \left. - \frac{r_{\rm h}^4b^2(E')}{j^4 \pi^4D^2(E')} \delta'(E-E')+... \right].
\end{eqnarray}
Here $r_{\rm h}$ is the radius of the region of interest.

Nevertheless, in the followings, we will analyze the X-ray data of $\omega$-Cen, which can be regarded as a very small dwarf galaxy. Many studies have suggested that $\omega$-Cen is quite likely the core of a captured and stripped dwarf galaxy \cite{Lee,Ibata}. For a small dwarf galaxy, the cooling effect of high-energy positrons and electrons is usually unimportant because its magnetic field strength and photon energy density (except the CMB component) is somewhat smaller than that of a normal galaxy. If the ICS of CMB is the dominated cooling process, we have $b(E)=2.5 \times 10^{-17}(E/1~{\rm GeV})^2$ GeV/s \cite{Vollmann}. For a high-energy electron produced from dark matter annihilation with energy $E \sim 1-1000$ GeV, its cooling timescale is $t_C \equiv E/b(E) \sim 10^{13}-10^{17}$ s while its diffusion timescale is $t_D \equiv r_{\rm h}^2/D(E) \sim 10^{11}-10^{12}$ s, with $r_{\rm h} \sim 5$ pc, $D_0 \sim 10^{26}$ cm$^2$/s and $\delta=0.3-0.7$. Therefore, we have $t_D/t_C \sim 10^{-5}-10^{-2} \ll 1$. The expanded terms inside the last bracket of Eq.~(4) for $j=1$ can be written as $\sim (t_D/\pi^2t_C)[1-(t_D/\pi^2t_C)+...]$. Since $t_D/\pi^2t_C<10^{-3}$, neglecting the second-order and higher-order expanded terms only contributes less than 1\% error in the analysis. Therefore, we only keep the leading expanded term for simplicity. Using the identity $\sum_{j=1}^{\infty} \sin(jy_1)\sin(jy_2) \equiv (\pi/2)\delta(y_1-y_2)$, we can get the solution of Eq.~(1) in the diffusion-dominated regime \cite{Vollmann}:
\begin{eqnarray}
\frac{dn_{\rm e}}{dE}(E,r)&=&\frac{1}{D(E)} \int_0^{r_{\rm h}}dr'\frac{r'}{r} \nonumber\\
&& \times \left[\frac12(r+r')-\frac12|r-r'|-\frac{rr'}{r_{\rm h}}\right]Q(E,r').
\label{dndE}
\end{eqnarray}  
In fact, the above solution can be obtained directly by neglecting the cooling term in Eq.~(1). In this regime, most of the high-energy electrons and positrons can completely diffuse out of the core region of the $\omega$-Cen without losing most of their energy. Since the core region we considered is less than 5 pc and the diffusion coefficient should be less than $10^{26}$ cm$^2$/s for $\omega$-Cen (see below), as mentioned above, we have $t_D \ll t_C$ (i.e. the diffusion rate is much larger than the cooling rate) and the diffusion-dominated regime solution can be applied. Since a smaller cooling rate $b(E)$ would give a smaller number of confined electrons and positrons inside a dwarf galaxy (i.e. a smaller value of $dn_{\rm e}/dE$), neglecting the cooling effect would suppress the ICS signal due to dark matter annihilation and the resulting constraints of the dark matter parameters obtained would be more conservative.

The high-energy electrons and positrons produced from dark matter annihilation would scatter with the CMB photons and boost the photon energy from $10^{-4}$ eV to $\sim$ keV through ICS. The number of CMB photons scattered per second from original frequency $\nu_0$ to new frequency $\nu$ is given by \cite{Blumenthal}
\begin{eqnarray}
I(\nu,x)&=&\frac{3\sigma_{\rm T}cn(\nu_0)}{16\gamma^4}\frac{\nu}{\nu_0} \nonumber\\
&&\times \left[2\frac{\nu}{\nu_0}\ln\left(\frac{1}{4\gamma^2}\frac{\nu}{\nu_0}\right)+\frac{\nu}{\nu_0}+4\gamma^2-\frac{1}{2\gamma^2}\frac{\nu^2}{\nu_0^2}\right],
\end{eqnarray}
where $\sigma_{\rm T}=6.65\times10^{-25}\,{\rm cm}^{2}$ is the Thomson cross section and $n(x)=170x^2/(e^x-1)\,{\rm cm}^{-3}$ is the number density of the CMB photons with $x=h\nu_0/kT_{\rm CMB}$, where $T_{\rm CMB}=2.725$ K is the present CMB temperature. The total X-ray luminosity due to the ICS can be given by \cite{Chan5}
\begin{eqnarray}
L_{\rm x}(m_{\rm DM})&=&2\int_0^{r_{\rm h}}dr\int_{m_e}^{m_{\rm DM}}dE\int_{E_1}^{E_2}d(h\nu) \int_0^{\infty}dx\nonumber\\
&&\times\left[\frac{dn_{\rm e}}{dE}(E,r)I(\nu,x)4\pi r^2\right]
\end{eqnarray}
where $E_1$ and $E_2$ are the lower and upper limits of the observed X-ray energy band. The factor two in Eq.~(7) indicates the contributions of both high-energy electrons and positrons. There are four parameters in the expression of $L_{\rm x}$: $\langle \sigma v \rangle$, $m_{\rm DM}$, $D_0$ and $\delta$. Therefore, the X-ray data can constrain the annihilating dark matter parameters $\langle \sigma v \rangle$ and $m_{\rm DM}$ for different sets of $D_0$ and $\delta$. 

Generally speaking, $D_0$ depends on the scale of structures and there is a physical range for galaxies. Theoretical model suggests that $D_0 \sim LV$ \cite{Rebusco,Jeltema,Heesen}, where $L$ is the injection scale and $V$ is the turbulent velocity. In our Milky Way, since $L \sim 1$ kpc and $V \sim 100$ km/s, we have $D_0 \sim 10^{28}$ cm$^2$/s. The actual values of $D_0$ modeled by early cosmic-ray studies give $D_0 \approx (0.1-3)\times 10^{28}$ cm$^2$/s \cite{Delahaye}. A more recent study has yielded a narrower range of $D_0$ for the benchmark cosmic-ray transport models: $D_0 \approx (0.6-3) \times 10^{28}$ cm$^2$/s \cite{Genolini}, which gives a good agreement with the theoretical model's prediction. Since the size of the region of interest in our study is only 3.9 pc (see below) and the dark matter density in $\omega$-Cen drops significantly outside 10 pc \cite{Brown,Evans}, we can conservatively set the injection scale $L$ to be less than 10 pc. Also, for $\omega$-Cen, the maximum velocity dispersion is smaller than 25 km/s \cite{Brown,Evans}. The turbulent velocity $V$ should also be less than 25 km/s. Therefore, we get a limit $D_0 \approx 8 \times 10^{25}$ cm$^2$/s. This value is the same order of magnitude ($D_0 \sim 10^{26}$ cm$^2$/s) as those commonly assumed in the studies of the Milky Way dwarf galaxies, such as the Ursa Major II galaxy \cite{Natarajan} and the dwarf spheroidal galaxies \cite{Jeltema,Kar}. Since the size of $\omega$-Cen ($\sim 10$ pc) is much smaller than the size of a typical dwarf galaxy ($\sim 1$ kpc), taking the limit $D_0 \le 10^{26}$ cm$^2$/s would be a conservative choice. We will also consider a more conservative limit $D_0 \le 10^{27}$ cm$^2$/s in our analysis.

For the diffusion index $\delta$, there are two benchmark diffusion models which predict $\delta=1/3$ (Kolmogorov model) \cite{Kolmogorov} and $\delta=1/2$ (Kraichnan model) \cite{Kraichnan}. We will consider a wider range $\delta=0.3-0.7$ in our analysis. This range is also consistent with that obtained by the cosmic-ray analysis of the Milky Way \cite{Delahaye,Genolini}. In Fig.~1, we show the prediction of $L_{\rm x}$ against $m_{\rm DM}$ for $D_0=10^{26}$ cm$^2$/s, $\langle \sigma v \rangle=2.2 \times 10^{-26}$ cm$^3$ s$^{-1}$ and two different values of $\delta$ (assumed $E_1=0.5$ keV and $E_2=6.0$ keV). We can see that the total luminosity is insensitive to the values of $\delta$, except for the $e^+e^-$ and $\mu^+\mu^-$ channels in the small $m_{\rm DM}$ regime. 

\begin{figure}
\begin{center}
\includegraphics[width=80mm]{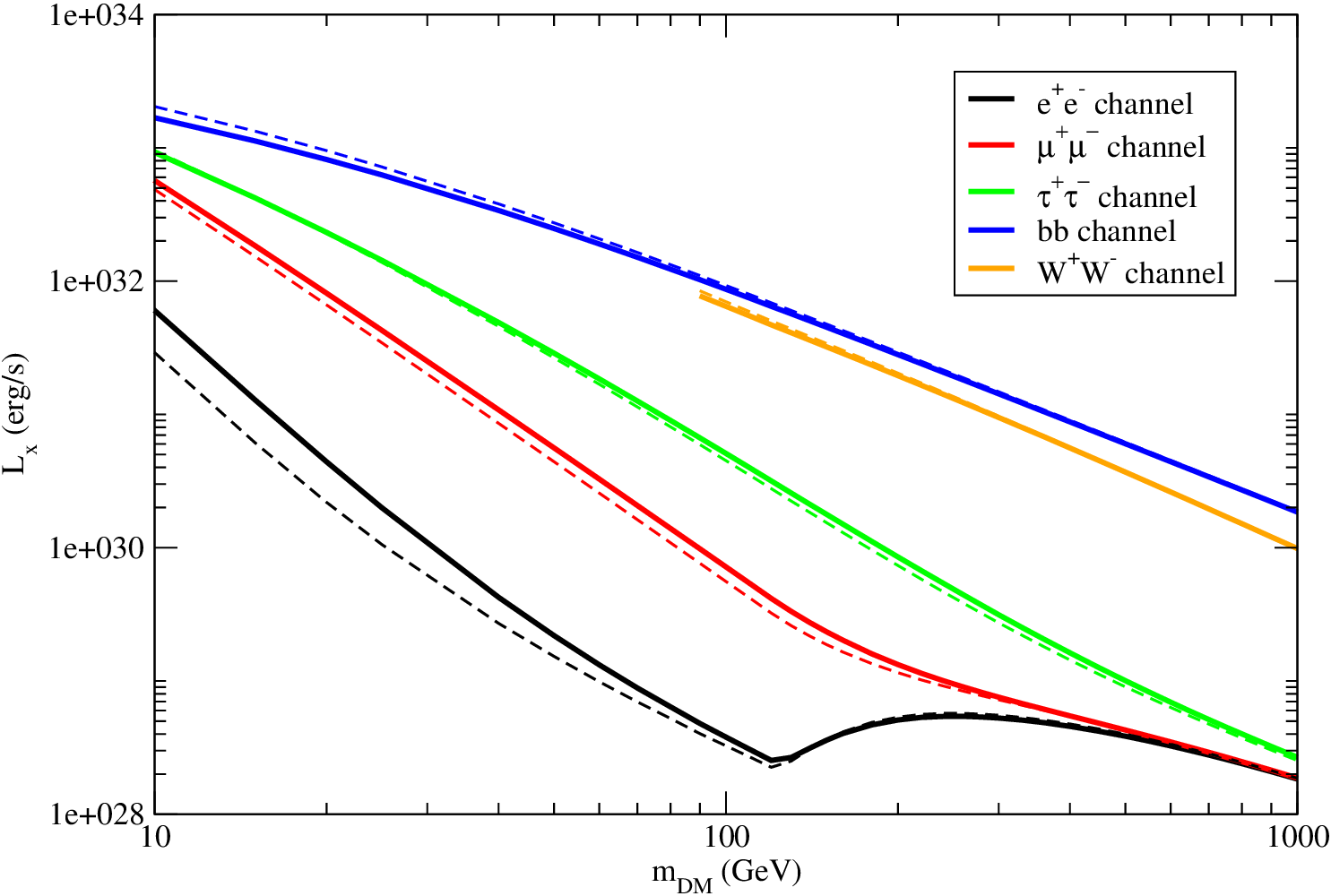}
\caption{The graph of $L_{\rm x}$ against $m_{\rm DM}$ for different annihilation channels with $\delta=0.3$ (solid lines) and $\delta=0.7$ (dashed lines). Here, we have assumed $D_0=10^{26}$ cm$^2$/s, $\langle \sigma v \rangle=2.2\times 10^{-26}$ cm$^3$/s, and the mean values of the dark matter density parameters $\rho_s=1.61\times 10^6M_{\odot}/{\rm pc}^3$ and $r_s=0.15$ pc (following the NFW profile).}
\label{Fig1}
\end{center}
\end{figure}

\section{Data of the $\omega$-Cen}
Recent studies of kinematic data show that the existence of dark matter in $\omega$-Cen is strongly statistically preferred relative to a stellar mass-only model \cite{Brown,Evans}. As many previous and recent studies have suggested that $\omega$-Cen is probably a captured and stripped dwarf galaxy \cite{Lee,Ibata}, the existence of dark matter in $\omega$-Cen is quite likely and its dark matter content might be very large. A more recent study has shown that the dark matter in $\omega$-Cen is much more centrally-concentrated so that it can give more stringent constraints for annihilating dark matter \cite{Evans}. 

We assume that the dark matter density profile of the $\omega$-Cen is spherically symmetric. The dark matter density profile can be best-described by the Navarro-Frenk-White (NFW) profile \cite{Navarro}:
\begin{eqnarray}
\rho_{\rm DM}(r)&=&
\frac{\rho_{\rm s}}{\frac{r}{r_s}\left(1+\frac{r}{r_{\rm s}}\right)^2}
\end{eqnarray}
where $r_s$ and $\rho_s$ are the scale radius and the scale density respectively. Using the latest kinematic data presented in \cite{Evans}, we can obtain the best-fit values with $1\sigma$ bounds: $\rho_s=1.61^{+2.02}_{-1.09}\times 10^6M_{\odot}/{\rm pc}^3$ and $r_s=0.15^{+0.05}_{-0.02}$ pc. Furthermore, \cite{Brown,Evans} show that another popular model of dark matter density profile, the Burkert profile \cite{Burkert}, does not have large difference in modelling the dark matter density. The Burkert profile is given by
\begin{equation}
\rho_{\rm DM}(r)=\frac{\rho_s}{\left(1+\frac{r^2}{r_{\rm s}^2}\right) \left(1+\frac{r}{r_{\rm s}} \right)}.
\end{equation}
The best-fit parameters with $1\sigma$ bounds are $\rho_s=4.10^{+4.80}_{-2.94} \times 10^6M_{\odot}/{\rm pc}^3$ and $r_s=0.10^{+0.05}_{-0.01}$ pc. In the followings, we will mainly follow the NFW profile to describe the dark matter distribution and we will consider the entire $1\sigma$ uncertainties of $\rho_s$ and $r_s$ in our analysis. As a reference, we will also show the corresponding results by following the Burkert profile. In Fig.~2, we show the variation of $L_{\rm x}$ against $m_{\rm DM}$ for different annihilation channels using the mean values, $1\sigma$ upper limits and $1\sigma$ lower limits of $\rho_s$ and $r_s$. Generally speaking, the $1\sigma$ lower limits of $\rho_s$ and $r_s$ can give the most conservative constraints of dark matter. Therefore, we will use the $1\sigma$ lower limits of $\rho_s$ and $r_s$ (i.e. the minimum dark matter contribution) to calculate the conservative bounds of $\langle \sigma v \rangle$ and $m_{\rm DM}$. 

On the other hand, $\omega$-Cen has been observed by the {\it Chandra X-ray Observatory} recently \cite{Henleywillis}. In the observation, 233 X-ray sources are identified and many of them are cataclysmic variables and their candidates. Also, those unidentified sources have luminosities and X-ray colors close to those of millisecond pulsars found in other globular clusters, and no abnormal case has been reported. Based on the X-ray analysis, the unabsorbed X-ray luminosity of $\omega$-Cen is $L_{\rm x} \le 1 \times 10^{30}$ erg s$^{-1}$ (with energy band $E=0.5-6.0$ keV and distance $=5.2$ kpc) for the region within the core radius $r=3.9$ pc \cite{Henleywillis}. Putting this upper limit into Eq.~(7) with $r_{\rm h}=3.9$ pc, $E_1=0.5$ keV and $E_2=6.0$ keV, we can get the constraints of dark matter mass for different annihilation channels. Here, in our first analysis, we leave the annihilation cross section $\langle \sigma v \rangle$ and the diffusion coefficient $D_0$ as free parameters. Since $L_{\rm x}$ is insensitive to the value of $\delta$, we simply fix $\delta=0.7$ and we can get the upper limits of $\langle \sigma v \rangle/D_0$ as a function of $m_{\rm DM}$ for each annihilation channel (see Fig.~3). Moreover, we can see in Fig.~3 that the difference in the upper limits between two dark matter density models (the NFW profile and the Burkert profile) is small.   

Generally speaking, the X-ray luminosity upper limit taken may contain some other background astrophysical emissions. If this is the case, the actual dark matter contribution on the X-ray luminosity would be smaller and the allowed dark matter parameter space would be more stringent. Therefore, neglecting the possible background astrophysical emissions can provide more conservative constraints on the dark matter parameters. 

If we take the upper limit $D_0 \le 10^{26}$ cm$^2$/s for $\omega$-Cen, we can obtain the upper limits of the annihilation cross section against $m_{\rm DM}$ for different annihilation channels (see Fig.~4 for the NFW profile and Fig.~5 for the comparison between two dark matter density models). Compared with the gamma-ray analyses of nearby dwarf galaxies from previous studies \cite{Ackermann,Albert}, we can see that our upper limits of $\langle \sigma v \rangle$ for non-leptophilic channels (e.g. $b\bar{b}$ and $W^+W^-$) are much more stringent than that obtained from gamma-ray analyses (see Fig.~4). For the leptophilic channels, our constraints are more stringent only in the small $m_{\rm DM}$ regime ($\le 100$ GeV). Specifically, if we take the thermal annihilation cross section $\langle \sigma v \rangle=2.2 \times 10^{-26}$ cm$^3$ s$^{-1}$ derived in standard cosmology \cite{Steigman}, we can get the lower limits of $m_{\rm DM}$ for thermal annihilating dark matter (see Table 1). These limits are more stringent than that obtained from gamma-ray analyses. We also show the lower limits of $m_{\rm DM}$ for $D_0 \le 10^{27}$ cm$^2$/s in Table 1 for comparison. The limits for the non-leptophilic channels are still more stringent than the gamma-ray limits.

\section{Discussion and conclusion}
In this article, by following the recent suggestions of the dark matter existence in $\omega$-Cen, we analyze the X-ray data of $\omega$-Cen to constrain annihilating dark matter. In particular, the conservative lower limits of thermal annihilating dark matter mass (assuming the NFW profile) can be constrained to 104(43) GeV, 650(167) GeV and 480(137) GeV for $\tau^+\tau^-$, $b\bar{b}$ and $W^+W^-$ channels respectively, assuming the diffusion coefficient $D_0 \le 10^{26}(10^{27})$ cm$^2$/s. The results will be slightly more conservative (less stringent) if we assume the Burkert profile to describe the dark matter density. Generally speaking, our results give more stringent constraints compared with the radio analysis \cite{Chan6,Vollmann2,Regis2}, gamma-ray analysis \cite{Ackermann,Albert} and neutrino analysis \cite{Antares}. These show that using X-ray data of appropriate dwarf galaxies might be able to give excellent constraints for annihilating dark matter. In particular, one recent radio analysis of the Large Magellanic Cloud has obtained a very robust stringent constraint $m_{\rm DM}<480$ GeV for quark annihilation channels, with the thermal annihilation cross section \cite{Regis2}. Our constraint on the $b\bar{b}$ channel for $D \le 10^{26}$ cm$^2$/s basically supports the stringent constraint obtained in \cite{Regis2} for the sub-TeV annihilating dark matter. 

In our analysis, two uncertain diffusion parameters ($D_0$ and $\delta$) are involved. For the diffusion index $\delta$, some models have predicted the possible range of $\delta$ \cite{Kolmogorov,Kraichnan} and we have examined a wider range $\delta=0.3-0.7$ to minimize the systematic uncertainty. For the diffusion coefficient $D_0$, theory can predict its order of magnitude \cite{Rebusco,Jeltema,Heesen} and the prediction is consistent with the observed range of $D_0$ in the Milky Way \cite{Delahaye,Genolini}. Based on the theoretical prediction, we expect that the value of $D_0$ for $\omega$-Cen should be $\le 10^{26}$ cm$^2$/s. Many studies have also assumed $D_0 \sim 10^{26}$ cm$^2$/s for dwarf galaxies \cite{Jeltema,Natarajan,Kar}. Since the value of $D_0$ depends on structural size, and the size of $\omega$-Cen is much smaller than that of a typical dwarf galaxy, we believe that taking $D_0 \le 10^{26}$ cm$^2$/s would be a conservative choice in our analysis. Even if we take a larger limit $D_0 \le 10^{27}$ cm$^2$/s, the lower limits of $m_{\rm DM}$ for the non-leptophilic channels are still more stringent than the gamma-ray limits (see Table 1). In fact, it is also possible to set $D_0$ to be a free parameter. In principle, we can determine the limits of dark matter mass in terms of $D_0$ by using Fig.~3 (assuming the thermal annihilation cross section). This is similar to the radio study of $\omega$-Cen and dwarf galaxies in \cite{Dutta}, which provides the limits of dark matter mass in terms of the parameter space of $D_0-\langle \sigma v \rangle$. Moreover, due to the small diffusion coefficient of $\omega$-Cen, the electrons and positrons can travel by a very long distance so that the actual diffusion halo can be larger than the size of $\omega$-Cen. Nevertheless, the ICS region that we considered is assumed to be the same region of the X-ray observations with radius $r_{\rm h}=3.9$ pc (see Eq.~(7)). In other words, the ICS luminosity outside the X-ray observation region has not been counted in our analysis.

Some previous gamma-ray studies of $\omega$-Cen have claimed the discovery of positive signals of annihilating dark matter \cite{Brown,Reynoso}. In the analysis of \cite{Brown}, the best-fit mass and cross section are $m_{\rm DM}=31\pm 4$ GeV and $\log (\langle \sigma v \rangle/{\rm cm^3s^{-1}})=-28.2^{+0.6}_{-1.2}$ respectively, annihilating via $b\bar{b}$ channel. In \cite{Reynoso}, two more possibilities have been suggested: $m_{\rm DM}=9.10^{+0.69}_{-0.62}$ GeV with $\log (\langle \sigma v \rangle/{\rm cm^3s^{-1}})=-26.5 \pm 0.03$ ($q\bar{q}$ channel) or $m_{\rm DM}=4.30^{+0.09}_{-0.08}$ GeV with $\log (\langle \sigma v \rangle/{\rm cm^3s^{-1}})=-25.34 \pm 0.03$ ($\mu^+\mu^-$ channel). Based on our analysis, only the suggestion in \cite{Brown} can escape from our stringent constraints. Our results can safely rule out the two possibilities suggested in \cite{Reynoso}. Furthermore, recent claims of dark matter interpretations of the gamma-ray excess and anti-proton excess in our Galaxy suggest $m_{\rm DM} \approx 31-40$ GeV with $\langle \sigma v \rangle \approx (1.4-2.0)\times 10^{-26}$ cm$^3$/s \cite{Daylan} and $m_{\rm DM} \approx 64-88$ GeV with $\langle \sigma v \rangle \approx (0.8-5.2)\times 10^{-26}$ cm$^3$/s \cite{Cholis} respectively, annihilating via $b\bar{b}$ channel. However, the constrained cross sections are larger than our conservative upper limits by more than an order of magnitude. Therefore, these claims are also safely ruled out. 

In our analysis, we have considered the s-wave dark matter annihilation (constant annihilation cross section) only. It is because the s-wave dark matter annihilation is the simplest model and the thermal annihilation cross section can be determined based on the standard cosmological model. However, if the s-wave annihilation is suppressed so that the p-wave annihilation is dominated, the annihilation cross section would be proportional to the velocity dispersion of dark matter \cite{Yang}. In the $\omega$-Cen, the velocity dispersion observed is relatively small ($<25$ km/s) so that the p-wave annihilation signal would not be very significant. However, if there exists a force carrier mediating the dark matter particle interaction, the annihilation cross section might be inversely proportional to the velocity dispersion of dark matter. This is known as the Sommerfeld enhancement \cite{Sommerfeld}. Due to the small velocity dispersion, the Sommerfeld enhancement might give a relatively larger annihilation signal in $\omega$-Cen. The parameters involved could be constrained by the observational data (e.g. gamma-ray data).

As the CMB spectrum is well-determined and the physics of ICS is well-understood, our analysis can give very stringent constraints on GeV to sub-TeV annihilating dark matter. It can provide a useful complementary analysis for constraining dark matter. In fact, it has been suggested for a long time that using X-ray data of dwarf galaxies can give certain constraints on GeV or sub-TeV annihilating dark matter \cite{Jeltema,Jeltema2,McDaniel2}. Nevertheless, many previous studies have mainly focused on the signals produced from dark matter with mass ranging from keV to sub-GeV (decaying dark matter or leptophilic annihilating dark matter) \cite{Bulbul,Boyarsky,Essig,Laha,Cirelli2}. Not many studies have practically applied our strategy to constrain dark matter with mass larger than GeV. Here we show that using X-ray data to constrain GeV to sub-TeV annihilating dark matter is very good, especially for the non-leptophilic channels. 

Moreover, using X-ray data to constrain dark matter has one large advantage compared with the gamma-ray analysis. The resolution of X-ray observations can be as small as $1"$ ($\sim$ keV) while the resolution of gamma-ray observations is usually larger than $5'$. Such a high resolution of X-ray detection can reveal the dark matter annihilation signal within a small central region of a dwarf galaxy. Since dark matter annihilation signals would be more prominent near the central region of a galaxy or a galaxy cluster due to the relatively high dark matter density, the observations using X-ray can focus on the small central emission region which has more dark matter potential contribution. Therefore, using X-ray data could be more likely to detect possible signal of dark matter annihilation or improve the lower limits of dark matter mass to a larger extent. We anticipate that future X-ray observations of small dwarf galaxies can possibly reveal the nature of dark matter. 

\begin{figure}
\begin{center}
\vskip 3mm
\includegraphics[width=80mm]{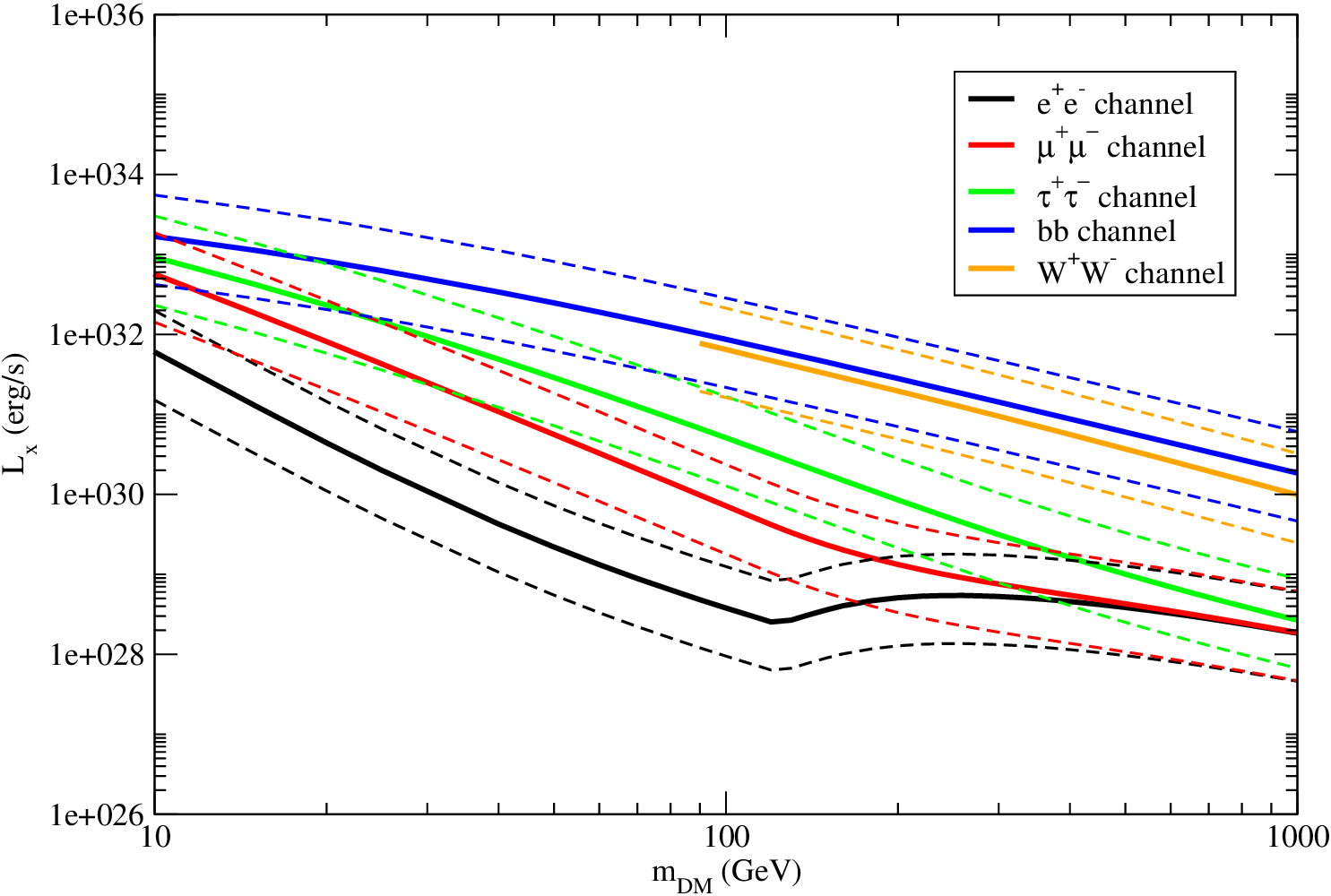}
\caption{The graph of $L_{\rm x}$ against $m_{\rm DM}$ for different annihilation channels with $\delta=0.3$. The solid lines represent the $L_{\rm x}$ with the mean values of the density parameters $\rho_s=1.61\times 10^6M_{\odot}/{\rm pc}^3$ and $r_s=0.15$ pc (following the NFW profile). The dashed lines represent the $L_{\rm x}$ calculated with the $1\sigma$ upper and lower limits of $\rho_s$ and $r_s$. Here, we have assumed $\langle \sigma v \rangle=2.2\times 10^{-26}$ cm$^3$/s and $D_0=10^{26}$ cm$^2$/s.}
\label{Fig2}
\end{center}
\end{figure}

\begin{figure}
\begin{center}
\vskip 3mm
\includegraphics[width=80mm]{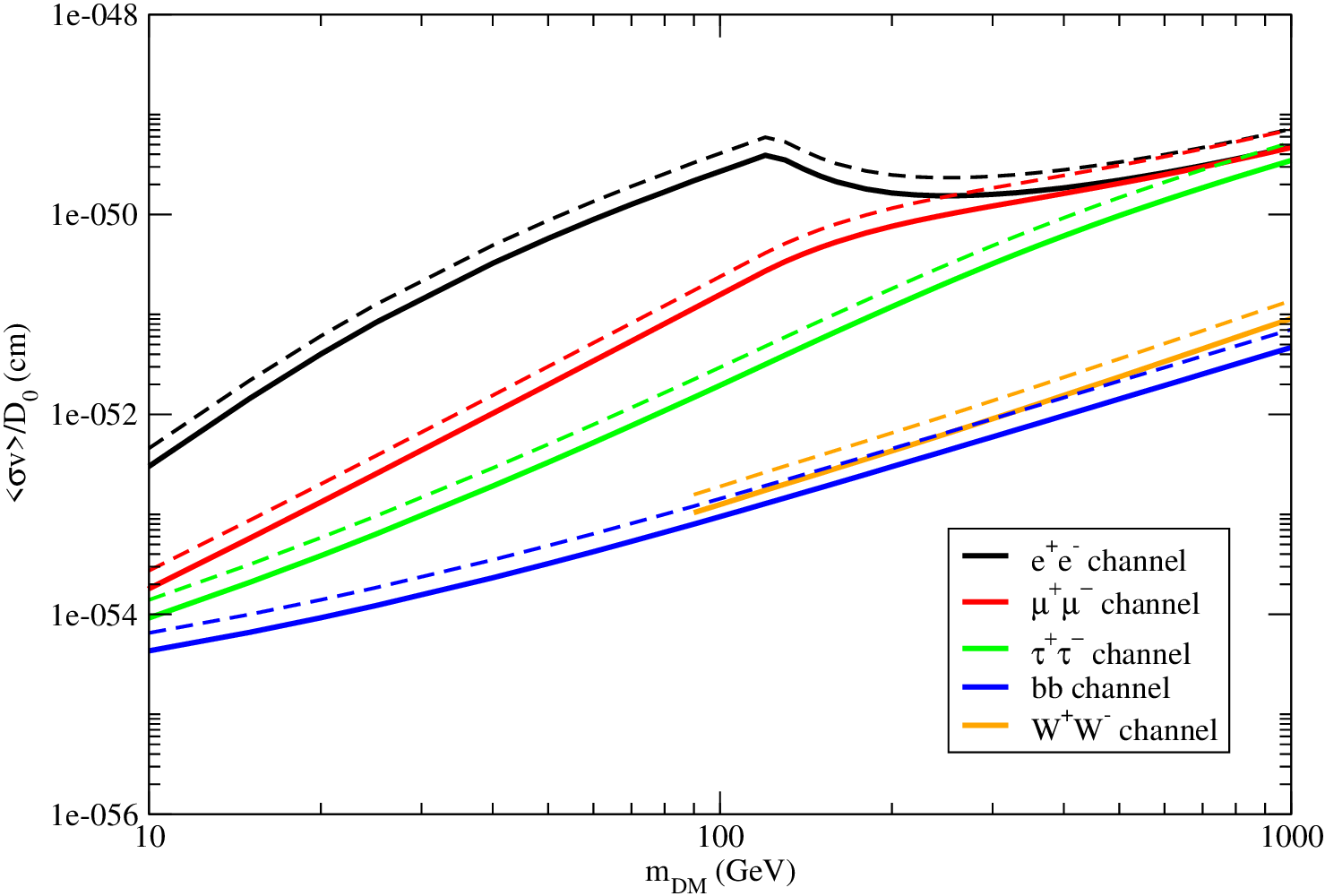}
\caption{The upper limits of $\langle \sigma v \rangle /D_0$ against $m_{\rm DM}$ with $\delta=0.7$ for different annihilation channels, taking $L_{\rm x} \le 1\times 10^{30}$ erg/s. The solid lines and dashed lines respectively represent the limits for the NFW profile and the Burkert profile. Here, we have taken the $1\sigma$ lower bounds of $\rho_s$ and $r_s$.}
\label{Fig3}
\end{center}
\end{figure}

\begin{figure}
\begin{center}
\vskip 3mm
\includegraphics[width=80mm]{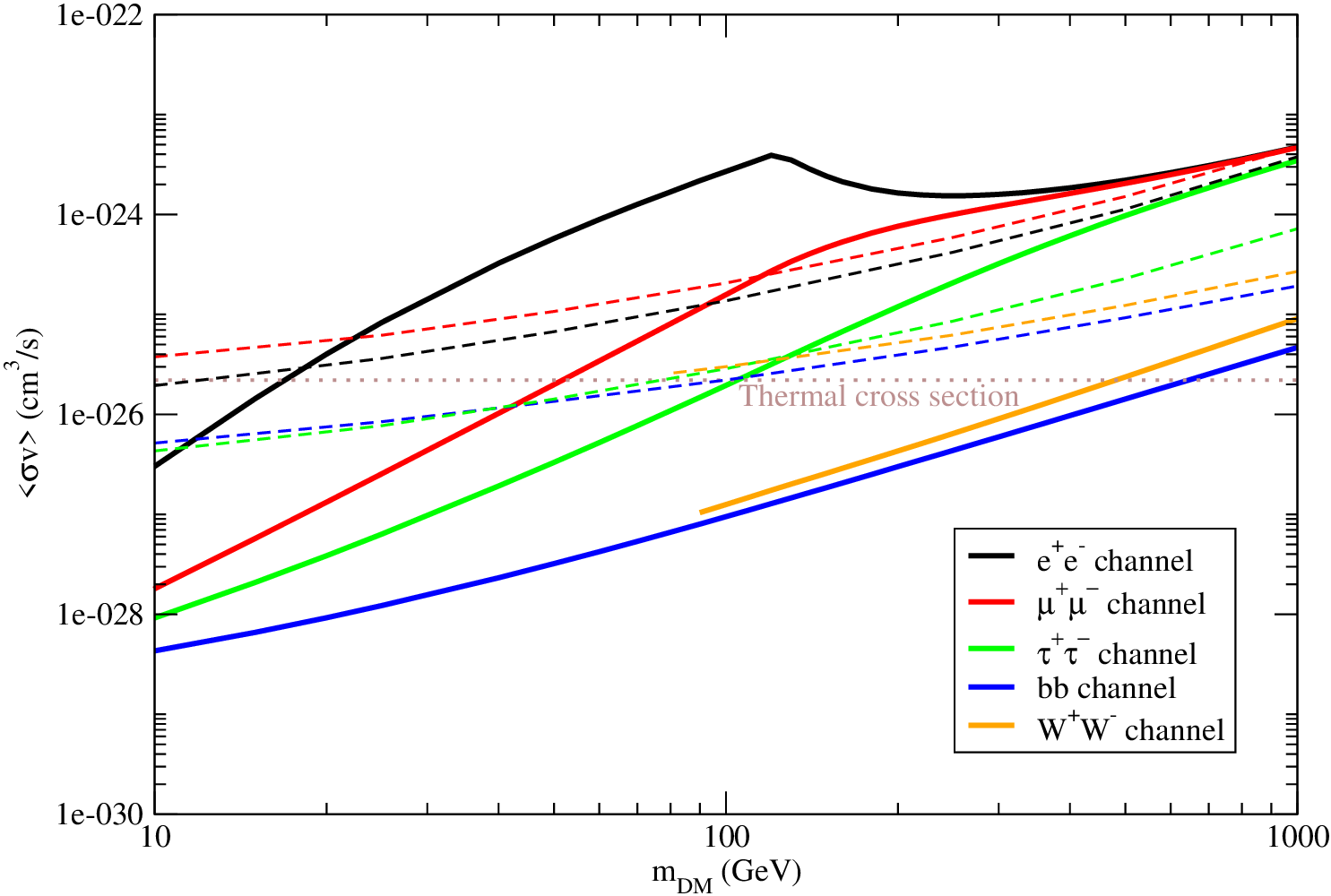}
\caption{The solid lines represent the upper limits of $\langle \sigma v \rangle$ against $m_{\rm DM}$ derived from our analysis with $\delta=0.7$ for different annihilation channels (assuming the NFW profile), taking $L_{\rm x} \le 1\times 10^{30}$ erg/s and $D_0=10^{26}$ cm$^2$/s. Here, we have assumed the $1\sigma$ lower bounds of $\rho_s$ and $r_s$. The dashed lines represent the upper limits of $\langle \sigma v \rangle$ derived from gamma-ray analysis of nearby dwarf galaxies \cite{Ackermann}. The brown dotted line indicates the thermal annihilation cross section $\langle \sigma v \rangle=2.2 \times 10^{-26}$ cm$^3$/s \cite{Steigman}.}
\label{Fig4}
\end{center}
\end{figure} 

\begin{figure}
\begin{center}
\vskip 3mm
\includegraphics[width=80mm]{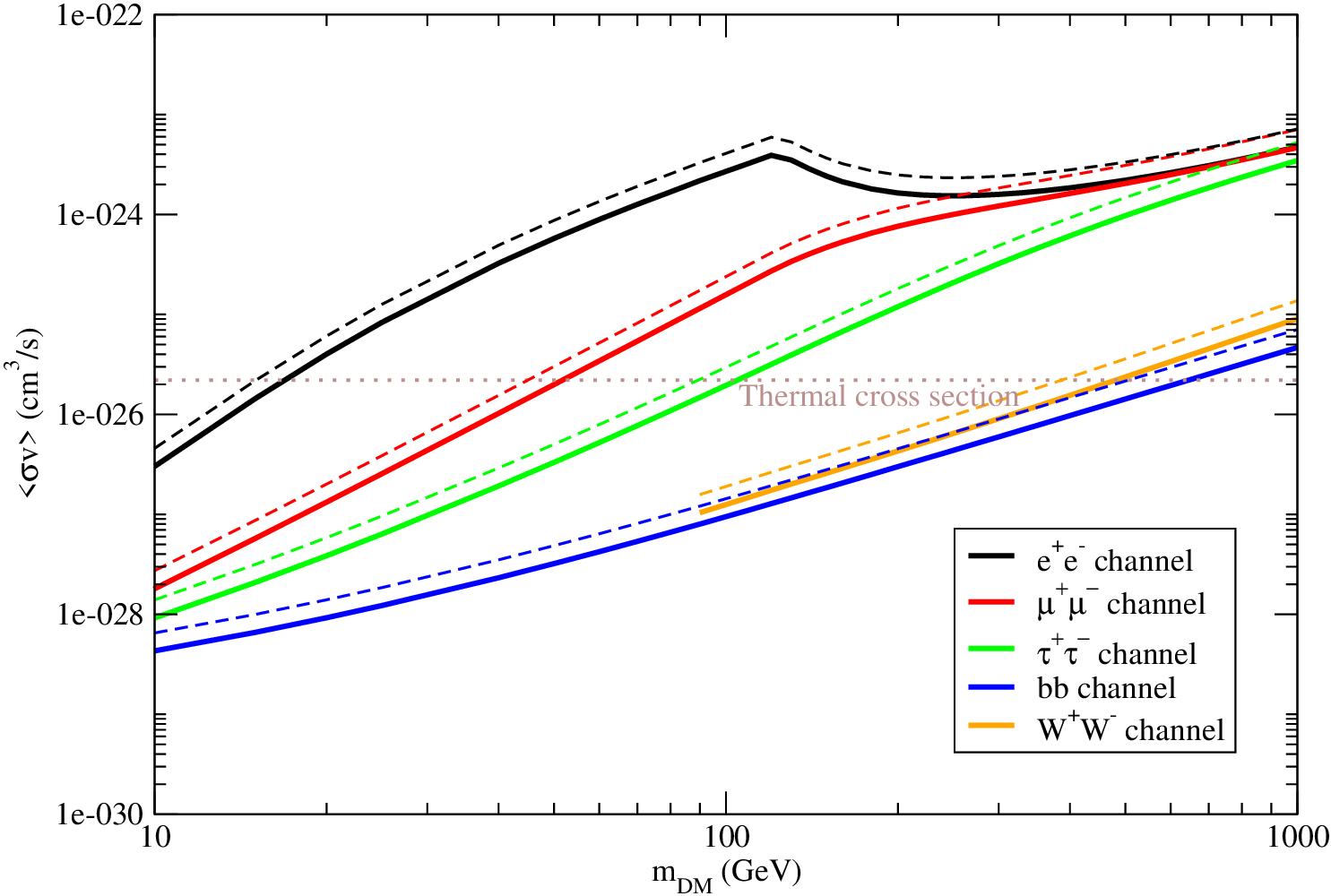}
\caption{The solid lines and dashed lines respectively represent the upper limits of $\langle \sigma v \rangle$ against $m_{\rm DM}$ derived from our analysis with $\delta=0.7$ for different annihilation channels following the NFW profile and the Burkert profile, taking $L_{\rm x} \le 1\times 10^{30}$ erg/s and $D_0=10^{26}$ cm$^2$/s. Here, we have assumed the $1\sigma$ lower bounds of $\rho_s$ and $r_s$. The brown dotted line indicates the thermal annihilation cross section $\langle \sigma v \rangle=2.2 \times 10^{-26}$ cm$^3$/s \cite{Steigman}.}
\label{Fig5}
\end{center}
\end{figure} 

\begin{table}
\caption{The lower limits of $m_{\rm DM}$ derived for thermal annihilating dark matter based on our analysis and gamma-ray analysis \cite{Ackermann}. The second and third columns are our results (following the NFW profile) while the fourth column are the results of gamma-ray analysis from \cite{Ackermann}. Here, we have assumed $\langle \sigma v \rangle=2.2 \times 10^{-26}$ cm$^3$/s and $\delta=0.7$.}
\begin{tabular}{lcccc}
 \hline\hline
Channel  & $m_{\rm DM}$ (GeV) & $m_{\rm DM}$ (GeV) & $m_{\rm DM}$ (GeV)  \\
         &  ($D_0 \le 10^{26}$ cm$^2$/s) & ($D_0 \le 10^{27}$ cm$^2$/s) &  Gamma-ray \\
\hline
$e^+e^-$ &  17 & 10 & 15 \\
$\mu^+\mu^-$ & 52 & 24 & 3 \\
$\tau^+\tau^-$ & 104 & 43 & 70 \\
$b\bar{b}$ & 650 & 167 & 100 \\
$W^+W^-$  & 480 & 137 & - \\
\hline
\label{table1}
\end{tabular}
\end{table}

\section{Acknowledgements}
We thank the anonymous referee for useful constructive feedbacks and comments. The work described in this paper was partially supported by the Seed Funding Grant (RG 68/2020-2021R) and the Dean's Research Fund of the Faculty of Liberal Arts and Social Sciences, The Education University of Hong Kong, Hong Kong Special Administrative Region, China (Project No.: FLASS/DRF 04628).

\label{lastpage}

\end{document}